\documentstyle[12pt]{article}

\newcommand{\Section}[1]{\section{#1}\setcounter{equation}{0}}
\renewcommand{\theequation}{\arabic{section}.\arabic{equation}}

         \def\ee{\end{equation}}
         \def\be{\begin{equation}}
         \def\bea{\begin{eqnarray}}
         \def\R{\rm {I\kern-.200em R}}
         \def\C{\rm {I\kern-.520em C}}
\begin{document}
\centerline{Revised Version}

\centerline{\bf Turbulent Two Dimensional Magnetohydrodynamics and Conformal
Field Theory}
\centerline{M. R. Rahimi Tabar}
\centerline{and}
\centerline{S. Rouhani}
\centerline{Department of Physics, Sharif University of Technology}
\centerline{Tehran P.O.Box: 11365-9161,Iran.}
\centerline{AND}
\centerline{Institute for Studies in Theoretical Physics and Mathematics}
\centerline{ Tehran P.O.Box: 19395-5746,Iran.}

\begin{abstract}
 We show that an infinite number of non-unitary minimal models may describe two
dimensional
turbulent magnetohydrodynamics (MHD), both in the presence and absence of the
Alf'ven effect.
 We argue that the existence of a critical dynamical index results in
the Alf'ven effect or equivelently the equipartition of energy. We show that
there are an
infinite number of conserved quantities in $2D-MHD$ turbulent systems both in
the limit of vanishing
the viscocities and in force free case.
In the force free case, using the non-unitary
minimal  model $ M_{2,7} $ we derive the correlation functions for the
velocity stream function and magnetic flux function. Generalising this simple
model we find the exponents of the energy spectrum in the inertial
range for a class of conformal field theories.
\end{abstract}
\newpage
\Section{- Introduction}
There has been some recent activity towards the application of Conformal Field
Theory (CFT) to the theory of turbulence in two dimensions [1-7]. The main
point is that the energy spectrum and higher correlation functins can be
derived by means of some non-unitary minimal
model of CFT .
Polyakov [1,2] derives a few criteria for a CFT which can be a possible
candidate for describing turbulence, and finds a candidate CFT which
gives the value of $ {{-25} \over 7} $
for the exponent of the energy spectrum (experimental results give an
exponent between 3 and 4 [11-14] ). Expanding on Polyakov's method
others [3,6] argue that there are a large number of CFT`s which satisfy
Polyakov's constraints but have more than one primary field. The role of
extra primary fields is not clear, but it has been suggested that
they may have to do with passive scalar and magnetic fields [4].
Briefly Polyakov's method is as follows. To describe turbulent behaviour
at high Reynold's numbers,  we interpret the Navier-Stokes equation
from a statistical mechanical point view.  That is we consider the
 correlation functions of the stream function with respect to some stationary
probability density:
$$ < \Phi _{i_1} (x_1) \Phi _{i_2} (x_2) ...\Phi _{i_N} (x_N)>.$$
as correlators of primary fields of some CFT .
The requirement that these correlations be stationary is the main prerequisite.
This condition
is refered to as the Hopf equation [9]. The fact that the correlation functions
have a power
law behaviour, indicates that conformal invariance may be at work . Indeed in
the limit of vanishing viscosity $ (\mu \rightarrow 0 ) $ the Navier-Stokes
equations
are scale invariant ( see [8] for more details ).
Also in a plasma one can use hydrodynamic equations to describe the collective
degrees of freedom, however the more conventional method uses the kinetic
equations to describe turbulent processes [10].
As in the case of pure hydrodynamics, the two dimensional magnetohydrodynamic
system (2D-MHD) differs from its three dimensional version in that it has an
infinite number of conserved quantities in the limit of zero viscocity
$ \mu \rightarrow 0 $ and zero molecular resistivity $\eta \rightarrow 0 $
and in this limit the equations are scale invariant.
Therefore 2D-MHD is also a good candidate where CFT may be applicable.
Such a system was first considered by Ferretti and Yang [4]. Here we shall
extend their arguments and find non-
unitary models which can give the correlation functions and the energy spectrum
index.
It is worth noting that the requirement that the critical dynamical index
be consistent, is equivalent to the Alf'ven effect, this means that there
is equipartion of energy between kinetic and magnetic coponents and this
requirement greatly reduces the number of possible minimal models of CFT
solutions. \\
This paper is organised as follows: in section 2 we first describe the
equations governing
 two dimensional magnetohydrodynamic systems (2D-MHD) in the inertial range.
 We then discuss its scaling properties. In section 3 we go on to describe a
conformal field theory
 with two primary fields pretaining to (2D-MHD).
 We then generalise to CFT's  with more primary fields and give a table of
 possible solutions in the general case and in the the Alf'ven region.
 In the appendix we describe a method of deriving OPE coefficients, for any
 Minimal Models in CFT and give the coefficients up to the third level.
The results are summarized in section 4 .
\Section{- Properties of two-dimensional MHD }
The incompressible two dimensional magnetohydrodynamic (2D-MHD)
 system has two independent dynamical variables, the velocity stream function
 $\varphi $, related to the velocity field $ V_\alpha $
\be V _\alpha = e _{\alpha \beta} \partial _{\beta} \varphi \ee
and the magnetic flux function  $ \psi$ related to the magnetic field $
B_\alpha $
 via:
\be B _ \alpha = e _ {\alpha \beta} \partial _{\beta} \psi \ee
here $ e _{\alpha \beta} $ is antisymmetric and $ e _{12}= +1 ,e_{21} =-1$. \\
The dynamics is given by the pair of equations $[10]$ :
\be {{\partial w} \over {\partial t}}= -e_{\alpha \beta }\partial _{\alpha}
\varphi \partial _{\beta} w + e_ {\alpha \beta} \partial _{\alpha} \psi
\partial _{\beta} J + \mu   \nabla^ 2 w \ee
\be {{\partial \psi} \over {\partial t}}= -e_{\alpha \beta}  \partial
_{\alpha} \varphi \partial_{\beta} \psi + \eta J \ee
 where
$$ w =\nabla ^ 2 \varphi  \hskip 2cm J=\nabla^ 2 \psi $$
here $\mu $ is the molecular kinematic viscosity and $\eta$ is the molecular
 resistivity. Note that normalisation is chosen such that the magnetic
 field assumes the same dimensions as velocity.
 In the inertial range $ \eta $ and  $\mu $ can be ignored then it follows
 from eqs. (2.3) and (2.4) that there exist three global, quadratic, conserved
 quantities.
\be E= {1\over 2} \int ( V^ 2 + B^ 2 ) d^ 2 x \ee
\be H= \int V.B d^2 x \ee
\be A= \int \psi^ 2 d ^2 x \ee
which are the total energy, the cross helicity and the mean square magnetic
 potential. In fact this system has an infinite number of conserved quantities
 such as:
\be R_n = \int \psi ^n d^2 x \ee
where n is any real number.
The time evolution of the E,H and A are given as:
\be {{dE}\over {dt}} = -\eta \int ( \nabla ^ 2 \psi ) ^ 2 d^ 2 x -\mu \int
 ( \nabla ^ 2 \phi ) ^ 2 d^ 2 x  \ee
\be {{dH} \over {dt}} =- ( \mu + \eta  ) \int  ( \nabla ^2 \psi ) ( \nabla^ 2
\varphi ) d^ 2 x \ee
\be {{dA} \over {dt}} = - \eta \int {( \nabla \psi )} ^ 2 d^ 2 x \ee
a similar expression for the $ R_n $ can be written as follows
\be {{dR_n} \over {dt}} = -\eta n ( n-1 ) \int \psi ^ {n-2} ( \nabla \psi ) ^ 2
d^ 2 x \ee
But when $ \eta $ and $ \mu $ are negligible, the system of equations (2.3) and
(2.4)
 display the scale invariance
\be  x \rightarrow \lambda x  \hskip 1cm \varphi \rightarrow \lambda^{1-h}
\varphi  \hskip 1cm  t \rightarrow \lambda ^{1+h}t  \hskip 1cm
\psi \rightarrow \lambda ^{1-h}\psi   \ee
Now if we impose Kolmogorov's idea of constant flux of energy we get $ h = {1
\over 3}$
[15], so that the behaviour of fields under scaling are :
\be  x \rightarrow \lambda x  \hskip 1cm  \varphi \rightarrow \lambda^ {2/3}
\varphi  \hskip 1cm
t \rightarrow \lambda ^ {4/3} t  \hskip 1cm   \psi \rightarrow \lambda ^ {2/3}
\psi \ee
Following Polyakov [24] we believe that this scale invariance signals
conformal symmetry of this system.
Simple scaling arguments [16-21], show that in turbulent 2D-MHD the energy
spectrum
behaves as:
 \be E ( k) \sim k^ {{-3} \over 2} \ee
Unfortunately this result is in poor agreement with recent numerical
simulations [29]. But our results show the deviation from this spectrum and is
in agreement with simulation [29].
 Another important aspect to consider is the Alf'ven effect [22-23]. This
effect
 essentially amounts to equipartition of energy between the kinetic and the
 magnetic components of the energy. Thus $ V^2_k $ and $ B^ 2_k $ should have
the same spectrum
 in the inertial range. We shall later show how this effect bears
 on the critical dynamical index, and limits our choice of CFT.\\
 If Polyakov's ideas were applicable here, a conformal field theory may exist
such that its correlation functions coincide with those of the 2D-MHD system.
However such a system has to be non-unitary in order to give the
power law behaviour suggested by the scaling relations (i.e. eq.(2.14)).
First of all, let us show that to describe turbulence by means of CFT ,
we have to use non-unitary minimal models.
Any local conformal field $ A_ j ( x) $ have associated with it an anamolous
dimension
$d _j $, i.e.
under a transformation $ x \rightarrow \lambda x $
we have [24,25] :
\be A _j (x ) \rightarrow \lambda ^ {-d_j} A _j (x ) \ee
We therefore observe that if associated with the fields of 2D-MHD system i.e.
the $\varphi$ and the $\psi$, they must have negative anamolous
dimensions. On the other hand negative  anamolous dimensions are possible
only in non-unitary minimal models which in turn non-unitary minimal models
result in
infrared divergences. We shall deal with this problem later.
\Section{- Conformal 2D-MHD Turbulence.}
The simplest model we consider is $ M_{2,7} $ with three primary fields I , the
identity
 $ \varphi $,  with anamolous dimension  $ ( {{-2} \over 7} ,{{-2} \over 7} )
$,
and  $ \psi $, with anamolous dimension  $ ( {{-3} \over 7} ,{{-3} \over 7} ) $
{}.
The central change is $ C ={{- 68} \over 7} $.
Now we can derive the small scale behaviour of the two point functions.
\be <\varphi_ {(x)} \varphi_ {(0)} > \sim |x| ^ {8 \over 7} \ee
\be < \psi _{(x)} \psi_ {(0)}> \sim |x| ^ {{12} \over 7}  \ee
and the correlation function of $ \varphi $ and $ \psi $ vanishes since they
have
different anamolous dimensions. However the above expressions  are clearly
unphysical since they grow with distance.
To avoid this problem one has to introduce an infrared cutoff [2].
Let us look at this problem in the momentum space
\be < \varphi _i (k )  \varphi _i ( -k ) > \sim C_i |k|^ {-2-4| \Delta \varphi
_i |} \ee
where $\varphi _i $ can be either $ \varphi$ or $\psi $ and
\be C_i=2^{4|\Delta \varphi _i|+ 1} {{ \Gamma (4 |\Delta \varphi_i|)} \over
{\Gamma(4|\Delta \varphi_i|+2 )}}  \ee
To avoid the problem of infrared divergence we can restrict ourselves to the
inertial range i.e. $ {1\over a} \gg K \gg {1 \over R} $ , where R is the large
scale boundary of the system and ${ a }$ is the dissipation range. Thus
\be <\varphi (x) \varphi (0)> \simeq C_i \int ^ \infty _
{k > {1 \over R}} k ^ {-2 -4| \Delta \varphi |} e^{{ik}.x}  dk \ee
which results in
\be <\varphi (x) \varphi (0) > \sim R ^ {4|\Delta \varphi_i |} -x^ {4 |\Delta
\varphi _i|}
+ \sum ^ {\infty}_ {m=1} (\alpha _m) ({{x} \over {2R}}) ^ m  x ^{4 |\Delta
\varphi_i |} \ee
where
\be \alpha_m= {{1}\over {m!}} {{\Gamma (4|\Delta \varphi_i|+1 )} \over { \Gamma
(2 |\Delta \varphi_i|+1
+{m \over 2})}} {{\sin ({1 \over 2} ( 1-4 |\Delta \varphi_i| -m ) \pi )} \over
{\Gamma ({1 \over 2} (4 |\Delta \varphi_i| + m +2 ) )}} \ee
Here the IR problem just as in the case of pure
turbulence [2,3,26],and it may be removed by considering of turbulent 2D-MHD
with boundary.\\
On the other hand, there exists a dissipation scale ${`a'}$
, so that the inertial
range was defined by $ k \ll {1\over a} $. Thus fully developed turbulence is
equivalent to letting $ `a' $ tend to zero, or equivalently letting the
Reynold's number tend to
infinity. However the existance of an ultraviolet cutoff means that we have
to be careful when  products of operators at the same point are involved.
To this end we handle such products using the point splitting technique.
Consider the right hand side of equation (2.3,2.4)
\begin{eqnarray}
 e _{\alpha \beta} \partial _{\alpha} \psi \partial _{\beta} \nabla^ 2 \psi=
\bar limit_{a \rightarrow 0}  e_{\alpha \beta} \partial _{\alpha} \psi (x+ {a
\over 2} ) \partial _{\beta} \nabla ^ 2 \psi ( x- {a \over 2} )
\end{eqnarray}
\begin{eqnarray}
e _{\alpha \beta} \partial _{\alpha} \varphi \partial_{\beta} \psi =
\bar limit_{a \rightarrow 0}  e_{\alpha \beta} \partial _{\alpha}
\varphi (x+ {a \over 2 }) \partial _{\beta} \psi ( x- {a\over 2} )
\end{eqnarray}
Where $\bar limit_{a \rightarrow 0} $, expresses angle averaging. Now to
evaluate
the above, we take
advantage of the operator product expansion. We have the general form for
the fusion rule of $( p,q) $ minimal model [27,28]:
\be [\psi_{{n_1}{ m_1}}] [\psi_ {{n_2},{m_2}}] = \sum ^{min
(n_1+n_2-1,2p-n_1-n_2-1)} _{k=|n_1 -n_2|+1}
\hskip 0.2cm \sum ^ {min(m_1+m_2-1,2q-m_1-m_2-1)} _{\ell =|m_1-m_2|+1 } [\psi
_{(k,\ell)} ] \ee
Where variables  $ k,\ell $ run over odd or even numbers if they are bounded by
odd or even numbers respectively. For our particular choice of
$(q,p)= (2,7 ) $ we have two primary fields  $\varphi$ and $\psi $ and their
families $ [\varphi]$ and $[\psi ]$ satisfy:
\be [\varphi ] \times [\varphi] = [I] + [ \psi] \ee
\be [ \varphi] \times [ \psi] = [ \psi ] + [ \varphi] \ee
\be [\psi] \times [ \psi ]= [ I] + [ \psi ]+ [ \varphi ] \ee
Note that by the family of $ \varphi $, we mean all operators which can be
constructed from $ \varphi $ using the Virasoro generators $ L_n $, such as:
\be L_{-n_1} L_{-n_2} ... L_{-n_k } \varphi  \ee
In order to explicitly calculate the rhs. of equations (3.8) and (3.9) by
means of the fusion rules, we use the following relations for the OPE of field
operators:
\be \varphi _n (x+ {a\over 2} ) \varphi _m ( x- {a\over 2} ) \sim |a| ^ { 2 (
\Delta_
p -\Delta_n - \Delta_ m )}
 ( \varphi _p + \alpha _1 aL_{-1} \varphi _p + a^ 2  {( \alpha _ 2 L_2 + \alpha
_3
L^ 2_ {-1} )} \varphi _p \ee
$$+ a^3 (\alpha _4 L^3_{-1} + \alpha_5 L_{-1} L_{-2}
+ \alpha_ 6 L_{-3} ) \varphi _p + ...$$
where $ n,m,p=1,2 $ and $\varphi _1 =\varphi , \varphi _2 = \psi $ . In the
appendix
we have given a method of deriving the coefficients $ \alpha _1 , \alpha_ 2,
\alpha _3 $
etc. and have derived these coefficients up to the third level.
By differentiation of lhs. of eq. (3.15) we will find the leading term in
product
of $ \varphi$ and $\psi $ in the limit $ a \rightarrow 0 $ as follows:
\be e_{ \alpha \beta } \partial _{\alpha} \varphi \partial _{\beta} \psi
= C_1 |a| ^ {{11} \over 7} ( \alpha  \bar{L} _{-1} ( L_{-2} + s L^ 2 _ {-1} )
-C.C ) \psi \ee
where s is a constant, determined by the operator product expansion and
$\alpha $ is a constant. It is clear that the antisymmetry of rhs. of
eq. (3.16) under complex conjugation is just the consequence of the $ e
$-tensor
at the lhs.
Similar calculation shows that the other terms in eqs. (2.3) and (2.4) by means
of OPE and point spliting procedure can be written in terms of Virasoro
generators in the $ limit$ $ a \rightarrow 0 $ as follows :
\be e_{\alpha \beta } \partial _{\alpha } \psi \partial _{\beta }
\partial ^ 2 \psi = C_2 |a| ^ {6 \over 7} {( L_{-2} \bar{L} ^ {-2} _{-1}
- \bar{L} _{2}  L^ 2 _{-1})} \psi \ee
\be e_{\alpha \beta } \partial _{\alpha} \varphi \partial _{\beta}
\partial ^ 2 \varphi = C_4 |a| ^ {1 \over 7} {( L_{-2} \bar{L}^ {-2}_{-1}
- \bar{L}_{2} L^ 2 _{-1})} \psi \ee
Before proceeding further let us look at the various constants of motion
$ R_n,A_2,H $and E :
\be {{dR_n} \over {dt}} = -\eta ( n-1 ) \int \psi ^ {n-1} \nabla^ 2
\psi d^ 2 X \ee
For $ R_n,A_2$,H and E to be constants of motion, we must have $ \mu,\eta = 0 $
$(\eta = {1 \over {\mu_{o} \sigma }} $ where $ \mu _0 $ and $ \sigma $ are
permeability and conductivity,
respectively) which require the conductivity $ \sigma $ to be infinite. On the
other
hand it is obvious that the above quantities are conserved in the limit of $
\mu$
and $ \nabla ^ 2 \psi \rightarrow 0 $, that is for finite conductivity
we have the same conserved quantities. But, letting $ \nabla ^2 \psi $ tend to
zero is equivalent to vanishing of
 $ J $. This situation is well known
as force free MHD [10].
Let us return to conformal field theory and look at $ \nabla ^ 2 \psi
\rightarrow 0$ , it is evident that;
\be \nabla ^ 2 \psi = 4 L_{-1} \bar{ L}_{-1}\psi =0 \ee
With this condition we can proceed further with the rhs. of equations (3.16),
(3.17), and (3.18) and we find that they are all zero.
In non- unitary minimal model $ M_{2,7} $ with $ C= {{-68} \over 7} $, $ \psi $
is degenarate
in the second level that is:
\be ( L_{-2}- {{21} \over 2} L^ 2 _{-1} ) \psi = 0, \Delta_{\psi} = {{-3} \over
7} \ee
Thus the rhs.s of eqs.(3.16),(3.17) and (3.18) vanish.
The more general $ N$-point  correlation function satisfies the well know BPZ
equation:
\be ( {{-21} \over 2} {{\partial ^ 2} \over {\partial z ^{2}}} - \sum ^ { n-1}
_ {j=1}
{{{-3} \over 7} \over {(z -z_j )}^ 2} + {{1} \over {( z- z_j )}} {{\partial}
\over
{\partial z_j}} )  < \psi (z_1) ... \psi ( z_{n-1}) > = 0 \ee
Now for a generelisation of our simple model let us postulate fusion rules as
below:
\be [\varphi] \times [\psi]= [X_1]+... \ee
where $ X_{1} $ , is the primary field of the lowest dimension in the OPE of $
\varphi$ and
$\psi $. Equation (3.16)  changes to
\be {{\partial \psi} \over {\partial t}}  \sim \lim  _ {a \rightarrow 0} |a| ^
{2 ( \Delta X - \Delta
\varphi - \Delta \psi ) }  ( \alpha \bar {L}_{-1} ( \beta L^ 2 _ {-1} + L_{-2}
)
-C.C. ) X_{1}  \ee
Let us now use the conserved quantities to find the constraints on dimensions
of the fields.
 Similar to the arguments used in [4],  we consider
the cascade of mean square magnetic potential:
\be A = \int \psi ^ 2  d ^ 2 x \ee
this  requires that $ < \dot{\psi} _{(x)} \psi _{(x)}> $be scale invariant.\\
Thus, using equations (3.24), we have
\be \Delta X_{1} + 2+ \Delta \psi = 0  \ee
\be \Delta X_{1} \geq \Delta \varphi + \Delta \psi \ee
where the second condition comes from  requiring non-singularity of the rhs.
of (3.24) in the limit $ a \rightarrow  0 $. Table 1 gives a list of models
which
satisfy these conditions. \\
Let us now consider the implications of the Alf'ven effect [22,23].
It is well known that [30] the excitations of the Elsasser's fields
propagate as the Alfv'en waves in opposite directions along the
lines of force of the `B' field at speeds of order `B'. What is meant by the
`inertial
range' in this context is, the region where the wavenumber of the the Alfv'en
waves lie whitin
the inertial range and the energy cascade results from the scattering of
Elssaser's fields [30].
This implies that , there should be asymptotically exact equipartition
of energy in the inertial range, i.e. $ {V_{k}}^2 = \alpha {B_{k}}^2 $ where $
\alpha $ is of order unity.
As discussed in detail by Chandrasekhar [31] , for Kolmogorov`s hypotheses  of
similarity
to holds here the constant $ \alpha $ shuld take value  1.62647 .
Therefore the equipartition
of energy  requires $ \varphi$  and $\psi $ to have similar scaling behavior:
\be \Delta \varphi = \Delta \psi \ee
We can derive the dynamical index of equations (2.3) and (2.4) as :
\be \Delta t = - \Delta \varphi + 2  \ee
and
\be \Delta t=\Delta \varphi -2\Delta \psi + 2 \ee
Where the eqs.(3.29) and (3.30) comes from eqs. (2.4) and (2.3) respectively.
Thus again the existence of a single index for temporal scaling require
$ \Delta \varphi = \Delta \psi $.
We can therefore extract models out of the table, which are consistent with the
Alf'ven effect. These are given in Table 2.
These models are consistent with the limit $ \mu \rightarrow  0 $ and $ \nabla
^ 2 \psi = 0 $, keeping $\sigma$, finite.\\
By means of the anomalous dimensions of $ \psi$ and $ \varphi $, we can
write the energy spectrum as follows:
$$ E(k) \sim k ^ {4| \Delta \varphi |+1 } + k^{ 4|\Delta \psi |+1 } $$
Where the exponents of energy spectrum for are given in Tables 1 and 2.
\newpage
\Section{- Concluding remarks}
In this paper we have derived a number of CFT which are possible
condidates for describing 2D-MHD. In the limit of
 finite conductivity requieres 2D-MHD to be force free and increases
the number of candidate CFT`s.  The imposition of the Alfv`en effect is
equivalent to
requiring a consistent  dynamical index and greatly reduces the number of
candidates.
A direct interpretation of primary fields is as yet, not possible, therefore
those candidate CFT's which have more than two primary fields seem as plausible
as
$ M_{2,7}$, which has two primary fields. Adding boundaries is the next
step in this theory, which may take it closer to some physical
problems such as the quantum wire.
Work along this line is already proceeding.\\
{\bf Acknowledgements. }
We are indebted to many of our colleagues for interesting and helpful
discussions, in particular M.Alimohmmadi, F.Ardalan, H.Arfaei,
S.Azakov and A.Yu.Morozov.
\newpage
{\bf Appendix:}
\renewcommand{\theequation}{A.\arabic{equation}}
{\bf Calculation of OPE Coefficients}\\
The most general expression for the operator product expansion is [27]:
\be \Phi_n (z,\bar{z} ) \Phi _m (o,o )= \sum _p \sum _{k} C^
{p,\{k\},\{\bar{k}\}} _{n m}
z^ {\Delta _p -\Delta _n - \Delta_m + \sum k_i } \bar{z} ^{ \bar{\Delta_p} -
\bar{\Delta_n}-\bar {\Delta_m}+\sum \bar{k_i}}  \Phi _p ^{\{k\},\{\bar k\}
}(0,0)  \ee %\eqnu  A-1
where the coefficients are
\be  C ^ { p,\{k\},\bar\{k\} } _{nm} = C ^{p}_{nm} \beta ^{ p, \{k\}}_{nm}
\bar{\beta} ^{p,\bar\{k\} }_ {nm}  \ee %\eqnu  A-2
\be  {\{k\}}={\{k_1 ,k_2,...,k_n \}}  \ee %\eqnu  A-3
Note that we have $ \varphi_i (0 ) |0>=| i> $, for the vacuum state $|0>$.
Now let equation $(A-1)$ act on $ |0> $:
\be\Phi_n (z ) | \Delta_m > = \sum C^ p _{nm} z ^ {\Delta_p - \Delta_n -
\Delta_m} \psi _p (z ) | \Delta_p > \ee %\eqnu  A-4
\be \psi_p (z) = \sum z ^ {\sum k_i} \beta ^ {p,\{K\}} _{nm} L_{ -k_1} ...
L_{ -k_n} \ee %\eqnu A-5
\be | z,\Delta_p > = \psi_ p|\Delta_p > \ee %\eqnu A-6 $$
Expand $ |z,\Delta_p> $ in terms of the complete basis $ |N,\Delta_p>$
where $| N,\Delta_p> $ is defined such that the coefficients of expansion are
$ z^N $:
\be |z,\Delta_p>=\sum z^N |N,\Delta_p> \ee %\eqnu A-7 $$
By applying of $ L_j $ over eq.(A-4) we have;
\be L_j|N+j,\Delta_p>= (\Delta_p-\Delta_m+j\Delta_n+N ) |N,\Delta_p> \ee
%\eqnu A-8 $$
and by solving the recursion relations we can find $ \beta ^{p,\{k\}}_{nm} $.
For level one we have:
\be L_1 |1,\Delta_p>=( \Delta_p-\Delta_m +\Delta_n ) | \Delta_p> \ee %\eqnu A-9
%%$$ \\
which results is
\be | 1,\Delta_p> = \alpha_1 L_{-1} |\Delta_p> \ee %\eqnu A-10 $$
or:
\be \alpha_1 = {{\Delta_p-\Delta_m+\Delta_n} \over {2\Delta_p }} \ee %\eqnu
%%A-11 $$ \\
For the second level by  means of eqs. (A-7) and (A-{10}) we have:
\be L_1 |2,\Delta_p> = {{(\Delta_p -\Delta _m + \Delta _n +1)(\Delta _p -
\Delta_m +\Delta_n )} \over {2 \Delta_p }} L_{-1} |\Delta_p > \ee %\eqnu A-12
%%$$
\be L_2 |2,\Delta_p> = ( \Delta_p-\Delta_m +2\Delta_n ) |\Delta_p >\ee%\eqnu
%%A-13 $$ \\
which results in:
\be |2,\Delta_p> = ( \alpha_2 L_{-2} + \alpha_ 3 L^ 2_{-1} ) | \Delta_p>\ee
%\eqnu  A-14 $$ \\
where $ \alpha_2$ and $\alpha_3 $ satisfy a system of equations:
$$ M_{ij} \alpha_j=A_i \hskip 2cm i,j=2,3 $$
where:\vskip 1.5cm
$$M=\left[ \begin{array}{cc}  4\Delta_p + {c/2}& 6\Delta_p\\
 O&4\Delta_p +2    \end{array}\right]  $$ \\ %\eqnu A-15 $$\\
\vskip 1.5cm
\be\left[ \begin{array}{c}  A_2 \\A_3 \end{array}\right]
=\left[\begin{array}{c} \Delta_ p -\Delta_ m+ 2\Delta_ n\\
 {{(\Delta_ p-\Delta_ m +\Delta_ n+1)(\Delta_ p-\Delta_ m +\Delta_ n)}\over
{2\Delta p}}
 \end{array} \right] \ee %\eqnu A-16 $$
This system can now be solved to give:
\be \left[ \begin{array}{c} \alpha_2 \\ \alpha_3 \end{array}\right]=
\left[\begin{array}{c}
6 \Delta_p A_3 - A_2 ( 4 \Delta_p+2 ) \\ 3A_1- A_2 (4 \Delta_p
+ {C \over 2} ) \end{array} \right] { 1 \over{ 2\Delta_p ( 5- 8 \Delta_p ) - (
2\Delta_p + 1)
 _C }} \ee %\eqnu A-17 $$
A similar method will work for higher levels. For level N, we have in place of
equation A-14, an expansion corresponding to the partition of N. We then find
a system of equations by successively applying $ L_j$, and finally the
coefficients are derived. For the third level, using equation A-8,  we have:
\be L_1|3,\Delta_p>=[\Delta_p-\Delta_m+2+\Delta_n]|2,\Delta_p> \ee % \eqnu A-18
%%$$
\be L_2|3,\Delta_p>=[\Delta_p-\Delta_m+1+2\Delta_n]|1,\Delta_p> \ee%  \eqnu
%%A-19 $$
\be  L_3|3,\Delta_p>=[\Delta_p-\Delta_m+3\Delta_n]|\Delta_p> \ee % \eqnu A-20
%%$$
 and $|3,\Delta_p> $ is given by:
\be |3,\Delta_p>=( \alpha_4 L^ 3_{-1} +\alpha_5 L_{-1} L_{-2}+\alpha_6 L_{-3}
) |\Delta_p> \ee %\eqnu A-21 $$
where $\alpha_4 $ ,$ \alpha_5 $ and $ \alpha_6 $  satisfy the following system
of
equations:
\be \alpha_4(24\Delta_p+6)+\alpha_5 \{{(4\Delta_p+{C\over 2})+9}\} +5\alpha_6=
{\acute{B} (\Delta_p+\Delta_n-\Delta_m)\over 2\Delta_p} \ee %\eqnu A-22 $$
\be \alpha_4(24\Delta_p)+\alpha_5(4(4\Delta_p+{c\over 2}))+
\alpha_6(6\Delta_p+2c) =\acute{C}  \ee %\eqnu A-23 $$
\be \alpha_4(8(\Delta_p+1))+\alpha_5 (7+2\Delta_p) +4 \alpha_6 =\acute{A}
(\alpha_2 +\alpha_3 )  \ee %\eqnu A-24 $$
and $\alpha_2,\alpha_3 $ are given by A-{15}. The coefficients $\acute{A},
\acute{B}$ and $\acute{C}$ are given by:
\be \acute{A}=[\Delta_p-\Delta_m+2+\Delta_n]   \ee %\eqnu A-25 $$
\be \acute{B}=[\Delta_p-\Delta_m+1+2\Delta_n]  \ee %\eqnu A-26 $$
\be \acute{C}=[\Delta_p-\Delta_m+3\Delta_n]  \ee %\eqnu A-27 $$
The inverse of the matrix of coefficients is :
\be M^{-1}_{ij} = {1 \over \Delta}  B_{ij} \ee %\eqnu A-28 $$
where\\
$B_{11}=C(2\Delta_p+3)+\Delta_p(6\Delta_p-11)$\\
$B_{12}=C+3\Delta_p+1/2$\\
$B_{13}=1/2[C^2+C(11\Delta_p+8)+2\Delta_p(12\Delta_p-13)]$\\
$B_{21}=4[C(\Delta_p+1)+3\Delta_p(\Delta_p-1)]$ \\
$B_{22}=2(7\Delta_p-2)$\\
$B_{23}=3[C(4\Delta_p+1)+\Delta_p(12\Delta_p-7)]$\\
$B_{31}=2[2C(\Delta_p+1)+5\Delta_p(2\Delta_p-1)]$\\
$B_{32}=2C(\Delta_p+1)-8\Delta^2_p-38\Delta_p+15$\\
$B_{33}=3[C(3\Delta_p+1)+2\Delta_p(12\Delta_p-5)]$\\
and
\be \Delta=2(2c^2(\Delta_p+1)+c(-2\Delta^ 2_p-16\Delta_p+7)-\Delta_p(24\Delta^
2_p -74\Delta_p+25)) \ee %\eqnu A-29 $$
\newpage
{\bf Table 1  Solution satisfying the constant mean square magnetic potential
flux condition }
\\
\[ \begin{array}{|c|c|c|c|c|c|} \hline(p,q) & \Phi & \Psi & X &
4{\Delta}_{\varphi} +1 & 4{\Delta}_{\psi} +1 \\ \hline
(2,13)&\Psi _{1,2}&\Psi _{1,4}&\Psi _{1,5}&-2.69&-.53
\\ \hline (2,17)&\Psi _{1,6}&\Psi _{1,2}&\Psi _{1,7}&-.64&-4.88
\\ \hline (2,19)&\Psi _{1,5}&\Psi _{1,2}&\Psi _{1,6}&-.68&-4.47
\\ \hline (2,23)&\Psi _{1,2}&\Psi _{1,3}&\Psi _{1,4}&-2.30&-.73
\\ \hline (2,27)&\Psi _{1,4}&\Psi _{1,2}&\Psi _{1,5}&-.77&-3.88
\\ \hline (3,29)&\Psi _{2,12}&\Psi _{1,3}&\Psi _{2,14}&-2.17&-1.20
\\ \hline (3,34)&\Psi _{1,2}&\Psi _{1,3}&\Psi _{1,4}&-2.29&-.73
\\ \hline (3,37)&\Psi _{1,5}&\Psi _{2,14}&\Psi _{2,18}&-.18&-5.05
\\ \hline (3,40)&\Psi _{1,4}&\Psi _{1,2}&\Psi _{1,5}&-.77&-3.87
\\ \hline (3,46)&\Psi _{2,17}&\Psi _{1,3}&\Psi _{2,19}&-2.47&-.21
\\ \hline (3,50)&\Psi _{1,3}&\Psi _{2,19}&\Psi _{2,21}&-1.4&-2.52
\\ \hline (3,52)&\Psi _{2,21}&\Psi _{1,2}&\Psi _{2,22}&-.82&-3.61
\\ \hline (3,62)&\Psi _{1,5}&\Psi _{2,22}&\Psi _{2,26}&-.37&-5.83
\\ \hline (3,77)&\Psi _{2,27}&\Psi _{1,3}&\Psi _{2,29}&-2.68&-.36
\\ \hline (3,79)&\Psi _{1,2}&\Psi _{2,29}&\Psi _{2,30}&-2.10&-.88
\\ \hline (3,80)&\Psi _{2,28}&\Psi _{1,3}&\Psi _{2,30}&-2.7&-.36
\\ \hline (3,82)&\Psi _{1,2}&\Psi _{2,30}&\Psi _{2,31}&-2.10&-.89
\\ \hline (3,89)&\Psi _{2,33}&\Psi _{1,2}&\Psi _{2,34}&-.89&-3.35
\\ \hline (3,92)&\Psi _{2,34}&\Psi _{1,2}&\Psi _{2,35}&-.90&-3.33
\\ \hline
\end{array}\]
\newpage
{\bf Table 2  Solutions that satisfy both the constant mean square magnetic
potential and the
Alf`ven constraint } \\  \\
\[ \begin{array}{|c|c|c|c|c|c|c|} \hline
(p,q)& \Phi & \Psi & X & 4{\Delta_\varphi+1}
& 4{\Delta_\psi+1}&\Delta t \\ \hline
 \hline (4,27)&\Psi_{2,10}&\Psi_{2,10}&\Psi_{3,19}&-2.08&-2.08&1.23
\\ \hline (5,32)&\Psi_{1,9}&\Psi_{1,9}&\Psi_{1,17}&-2.5&-2.5&1.12
\\ \hline (5,34)&\Psi_{2,10}&\Psi_{2,10}&\Psi_{3,19}&-2.04&-2.04&1.24
\\ \hline (6,35)&\Psi_{2,11}&\Psi_{2,12}&\Psi_{3,22}&-2.98&-2.92&1.0
\\ \hline (6,35)&\Psi_{3,17}&\Psi_{3,17}&\Psi_{5,33}&-2.96&-2.96&1.01
\\ \hline (6,37)&\Psi_{1,8}&\Psi_{1,8}&\Psi_{1,15}&-2.78&-2.78&1.06
\\ \hline (6,41)&\Psi_{2,10}&\Psi_{2,10}&\Psi_{3,19}&-2.01&-2.01&1.25
\\ \hline \end{array}\]

\end{document}